\documentclass[preprintnumbers,unsortedaddress,superscriptaddress,notitlepage]{revtex4}
\usepackage{amsfonts}
\usepackage{amsmath}
\usepackage{hyperref}
\usepackage{amssymb}
\usepackage[english]{babel}
\usepackage{graphicx}
\usepackage{epsfig}
\usepackage{bm}
\usepackage{longtable}
\usepackage{verbatim}
\usepackage{longtable}
\newcommand{\mathsym}[1]{{}}

\newcommand{\emp}{\begin{equation}}
\newcommand{\fin}{\end{equation}}
\newcommand{\empn}{\begin{equation*}}
\newcommand{\finn}{\end{equation*}}

\newcommand{\bea}{\begin{eqnarray}}
\newcommand{\eea}{\end{eqnarray}}
\newcommand{\eger}{\begin{gather}}
\newcommand{\fger}{\end{gather}}
\newcommand{\egn}{\begin{gather*}}
\newcommand{\fgn}{\end{gather*}}
\newcommand{\bit}{\begin{itemize}}
\newcommand{\eit}{\end{itemize}}

\input{tcilatex}
\begin{document}

\title{Lepton masses and mixings in an $A_{4}$ multi-Higgs model with radiative seesaw mechanism}
\author{A. E. C\'arcamo Hern\'andez}
\email{antonio.carcamo@usm.cl}
\affiliation{{\small Universidad T\'ecnica Federico Santa Mar\'{\i}a and
Centro Cient\'{\i}fico-Tecnol\'ogico de Valpara\'{\i}so}\\
Casilla 110-V, Valpara\'{\i}so, Chile}
\author{I. de Medeiros Varzielas}
\email{ivo.de@unibas.ch}
\affiliation{{\small Facult\"at f\"ur Physik, Technische Universit\"at Dortmund}\\
D-44221 Dortmund, Germany}
\affiliation{{\small Department of Physics, University of Basel,}\\
Klingelbergstr. 82, CH-4056 Basel, Switzerland}
\author{S. Kovalenko}
\email{sergey.kovalenko@usm.cl}
\affiliation{{\small Universidad T\'ecnica Federico Santa Mar\'{\i}a and Centro Cient%
\'{\i}fico-Tecnol\'ogico de Valpara\'{\i}so}\\
Casilla 110-V, Valpara\'{\i}so, Chile}
\author{H. P\"as}
\email{heinrich.paes@uni-dortmund.de}
\affiliation{{\small Facult\"at f\"ur Physik, Technische Universit\"at Dortmund}\\
D-44221 Dortmund, Germany}
\author{Iv\'an Schmidt}
\email{ivan.schmidt@usm.cl}
\affiliation{{\small Universidad T\'ecnica Federico Santa Mar\'{\i}a and Centro Cient%
\'{\i}fico-Tecnol\'ogico de Valpara\'{\i}so}\\
Casilla 110-V, Valpara\'{\i}so, Chile}

\begin{abstract}
We propose a renormalizable multi-Higgs model with $A_{4}\otimes Z_{2}\otimes Z^{\prime}_{2}$ symmetry,
accounting for the experimental deviation from the tribimaximal mixing pattern of the neutrino mixing matrix.
In this framework we study the charged lepton and neutrino masses and mixings.
The light neutrino masses are generated via a radiative seesaw mechanism, which involves a single heavy Majorana neutrino and neutral scalars running in the loops.
The obtained neutrino mixings and mass squared splittings are in
good agreement with the neutrino oscillation experimental data for both normal and inverted hierarchy.
The model predicts an effective Majorana neutrino mass $m_{\beta\beta}=$ 4 meV and 50 meV for the normal and the inverted neutrino spectrum, respectively. 
The model also features a suppression of CP violation in neutrino oscillations, a low scale for the heavy Majorana neutrino (few TeV) and, due to the unbroken $Z_2$ symmetry, a natural dark matter candidate.
\end{abstract}

\maketitle

\section{Introduction}

The existence of three generations of fermions, as well as their particular pattern of masses and mixing cannot be understood within the Standard Model (SM), and
makes it appealing to consider a more fundamental theory addressing these issues.
This problem is especially challenging in the neutrino sector, where
the striking smallness of neutrino masses and large mixing between generations suggest a different kind of underlying physics than what should be responsible for the masses and mixings of the quarks.
Unlike in the quark sector, where the mixing angles are very small, two of the three neutrino mixing angles,
the atmospheric $\theta_{23}$ and the solar $\theta_{12}$ are large, while the reactor angle $\theta_{13}$ is comparatively small
\cite{PDG,Abe:2011sj,Adamson:2011qu,Abe:2011fz,An:2012eh,Ahn:2012nd,Tortola:2012te,Fogli:2012ua,GonzalezGarcia:2012sz}.

In the literature there has been a formidable amount of effort to understand the origin of the leptonic flavor structure, with various proposed scenarios and models of neutrino mass generation.
Among those approaches to understand the pattern of neutrino mixing, models with discrete flavor symmetries are particularly popular 
(for recent reviews see Refs. \cite{King:2013eh,Altarelli:2010gt,Ishimori:2010au}).
There is a great variety of such models, some with Multi-Higgs sectors \cite{textures,Ma:2006km,Ma:2006fn,Hernandez:2013mcf,Machado:2010uc,Ma:2001dn,Babu:2002dz,Altarelli:2005yx,Altarelli:2009gn,Bazzocchi:2009pv,He:2006dk,Fukuyama:2010mz,Fukuyama:2010ff,Holthausen:2012wz,Ahn:2012tv,Ahn:2013mva,Chen:2012st,Toorop:2010ex,Memenga:2013vc,Ferreira:2013oga,Felipe:2013vwa,Machado:2007ng,Machado:2011gn,Felipe:2013ie,Varzielas:2012ai,Ishimori:2012fg,Bhattacharya:2013mpa,Teshima:2005bk,Mohapatra:2006pu,Ma:2013zca,Canales:2013cga,Canales:2012dr,Dong:2011vb,Kajiyama:2013sza,Hernandez:2013hea,Mohapatra:2012tb,Varzielas:2012pa,Ding:2013hpa,Cooper:2012bd,King:2013hj,Morisi:2013qna,Morisi:2013eca,Varzielas:2012nn, Bhattacharyya:2012pi,Ma:2013xqa,Nishi:2013jqa,Varzielas:2013sla,Aranda:2013gga},
Extra Dimensions
\cite{Rius:2001dd,Dobrescu:1998dg,Altarelli:2005yp,Ishimori:2010fs,Kadosh:2010rm,Kadosh:2013nra,Ding:2013eca,CarcamoHernandez:2012xy}, Grand Unification \cite{GUT} or Superstrings \cite{String}.
Another approach attempts to describe certain phenomenological features of the fermion mass hierarchy by postulating particular zero-texture Yukawa matrices \cite{textures}.

In this context, the groups explored recently in the literature include $A_4$ \cite{He:2006dk,Fukuyama:2010mz,Fukuyama:2010ff,Ahn:2012tv,Ahn:2013mva,Chen:2012st,Toorop:2010ex,Memenga:2013vc,Holthausen:2012wz,Ferreira:2013oga,Felipe:2013vwa,Machado:2007ng,Machado:2011gn,Felipe:2013ie,Varzielas:2012ai,Ishimori:2012fg,Bhattacharya:2013mpa,King:2013hj,Morisi:2013qna,Morisi:2013eca,Altarelli:2005yp,Kadosh:2010rm,Kadosh:2013nra},
$\Delta(27)$ \cite{Varzielas:2012nn,Bhattacharyya:2012pi,Ma:2013xqa,Nishi:2013jqa,Varzielas:2013sla,Aranda:2013gga}
, $S_3$ \cite{Teshima:2005bk,Mohapatra:2006pu,Ma:2013zca,Canales:2013cga,Canales:2012dr,Dong:2011vb,Kajiyama:2013sza,Hernandez:2013hea}, $S_4$ \cite{Altarelli:2009gn,Bazzocchi:2009pv,Mohapatra:2012tb,Varzielas:2012pa,Ding:2013hpa,Ding:2013eca,Ishimori:2010fs}
and $A_5$ \cite{Cooper:2012bd}.
These models can be implemented in a supersymmetric framework \cite{Babu:2002dz,Altarelli:2005yx,Altarelli:2009gn,Bazzocchi:2009pv,Morisi:2013eca}, or in extra dimensional scenarios with $S_4$ \cite{Ishimori:2010fs,Ding:2013eca} or $A_4$ \cite{Altarelli:2005yp,Kadosh:2010rm,Kadosh:2013nra}.

The popular tribimaximal (TBM) ansatz for the leptonic mixing matrix
\begin{equation}
\label{TBM-ansatz}
U_{TBM}=\left(
\begin{array}{ccc}
\sqrt{\frac{2}{3}} & \frac{1}{\sqrt{3}} & 0 \\
-\frac{1}{\sqrt{6}} & \frac{1}{\sqrt{3}} & -\frac{1}{\sqrt{2}} \\
-\frac{1}{\sqrt{6}} & \frac{1}{\sqrt{3}} & \frac{1}{\sqrt{2}}%
\end{array}%
\right) ,
\end{equation}
can originate, in particular, from $A_{4}$. TBM
corresponds to mixing angles with $\left(\sin ^{2}\theta_{12}\right) _{TBM}=\frac1{3}$,
$\left(\sin^{2}\theta _{23}\right)_{TBM}=\frac1{2}$, and
$\left(\sin ^{2}\theta_{13}\right)_{TBM}=0$.
On the other hand the T2K \cite{Abe:2011sj}, MINOS \cite{Adamson:2011qu}, Double Chooz \cite{Abe:2011fz}, Daya Bay \cite{An:2012eh} and
RENO \cite{Ahn:2012nd} experiments have recently measured a non-vanishing mixing angle $\theta_{13}$, ruling out the exact TBM pattern.
The global fits of the available data from neutrino oscillation experiments \cite{Tortola:2012te, Fogli:2012ua, GonzalezGarcia:2012sz} give experimental constraints on the neutrino mass squared splittings and mixing parameters. We use the values from \cite{Tortola:2012te}, shown in Tables \ref{NH} and  \ref{IH}, for the cases of normal and inverted hierarchy, respectively. It can be seen that the data deviate significantly from the TBM pattern.
\begin{table}[tbh]
\begin{tabular}{|c|c|c|c|c|c|}
\hline
Parameter & $\Delta m_{21}^{2}$($10^{-5}$eV$^2$) & $\Delta m_{31}^{2}$($%
10^{-3}$eV$^2$) & $\left( \sin ^{2}\theta _{12}\right) _{\exp }$ & $\left(
\sin ^{2}\theta _{23}\right) _{\exp }$ & $\left( \sin ^{2}\theta
_{13}\right) _{\exp }$ \\ \hline
Best fit & $7.62$ & $2.55$ & $0.320$ &  $0.613$ & $0.0246$ \\
\hline
$1\sigma $ range & $7.43-7.81$ & $2.46-2.61$ & $0.303-0.336$ & $0.573-0.635$ & $0.0218-0.0275$ \\ \hline
$2\sigma $ range & $7.27-8.01$ & $2.38-2.68$ & $0.29-0.35$ & $0.38-0.66$ & $%
0.019-0.030$ \\ \hline
$3\sigma $ range & $7.12-8.20$ & $2.31-2.74$ & $0.27-0.37$ & $0.36-0.68$ &
\\ \hline
\end{tabular}%
\caption{Range for experimental values of neutrino mass squared splittings
and leptonic mixing parameters taken from Ref.  \protect\cite{Tortola:2012te} for the case of normal hierarchy.}
\label{NH}
\end{table}
\begin{table}[tbh]
\begin{tabular}{|c|c|c|c|c|c|}
\hline
Parameter & $\Delta m_{21}^{2}$($10^{-5}$eV$^2$) & $\Delta m_{13}^{2}$($%
10^{-3}$eV$^2$) & $\left( \sin ^{2}\theta _{12}\right) _{\exp }$ & $\left(
\sin ^{2}\theta _{23}\right) _{\exp }$ & $\left( \sin ^{2}\theta_{13}\right) _{\exp }$ \\ \hline
Best fit & $7.62$ & $2.43$ & $0.320$ & $0.600$ & $0.0250$ \\ \hline
$1\sigma $ range & $7.43-7.81$ & $2.37-2.50$ & $0.303-0.336$ & $0.569-0.626$
& $0.0223-0.0276$ \\ \hline
$2\sigma $ range & $7.27-8.01$ & $2.29-2.58$ & $0.29-0.35$ & $0.39-0.65 $
& $0.020-0.030$ \\ \hline
$3\sigma $ range & $7.12-8.20$ & $2.21-2.64$ & $0.27-0.37$ & $0.37-0.67$ & $%
0.017-0.033$ \\ \hline
\end{tabular}%
\caption{Range for experimental values of neutrino mass squared splittings and leptonic mixing parameters taken from Ref.  \protect\cite{Tortola:2012te} for the case of inverted hierarchy.}
\label{IH}
\end{table}\newline

Here we present a renormalizable model with $A_{4}\otimes Z_{2}\otimes Z^{\prime}_{2}$ discrete flavor symmetry,
which is consistent with the current neutrino data for the neutrino masses and mixings shown in Tables \ref{NH},\ref{IH} and which has less effective model parameters than other similar models, as discussed in section \ref{Numerical Analysis}.
We choose $A_4$ since it is the smallest symmetry with one three-dimensional
and three distinct one-dimensional irreducible representations, where the three families of fermions can be accommodated rather naturally. Thereby we unify the left-handed leptons in the $A_4$ triplet representation and assign the right-handed leptons to $A_4$ singlets. This type of setup was proposed for the first time
in Ref. \cite{Ma:2001dn}.
In our model there is  only one right-handed SM singlet Majorana neutrino $N_{R}$,
and the scalar sector includes three $A_4$ triplets, one of which is a SM singlet while the other two are $SU(2)_L$ doublets.
We further impose on the model a $Z_2$ discrete symmetry, in order to separate the two $A_{4}$ triplets transforming as $SU(2)_{L}$ doublets, so that one of them
participates only in those Yukawa interactions which involve right-handed $SU(2)_{L}$   singlets $e_{R}, \mu_{R}, \tau_{R}$, while
the other one participates only in those with the right-handed SM sterile neutrino $N_{R}$.
Finally, a (spontaneously broken) $Z^{\prime}_2$ symmetry is introduced to forbid terms in the scalar potential with odd powers of the SM singlet scalar field $\chi$, the only one transforming non-trivially under $Z^{\prime}_2$.
We assume that the $Z_2$ symmetry  is not affected by the Electroweak Symmetry Breaking. Therefore the scalar fields coupled to the neutrinos
have vanishing vacuum expectation values, which implies that the light neutrino masses are not generated at tree level via the usual seesaw mechanism, but instead
are generated through loop corrections in a variant of the so-called radiative seesaw mechanism.
The loops involve a heavy Majorana neutrino and neutral scalars, which in turn couple through quartic interactions with other neutral scalars in the external lines.
The smallness of neutrino masses generated via a radiative seesaw mechanism is attributed  to the smallness of the loop factor and to the quadratic dependence on the small neutrino Yukawa coupling.
The scale of new physics can therefore be kept low, with the heavy Majorana neutrino mass of a few TeV.
The radiative seesaw mechanism has been discussed in Refs. \cite{Ahn:2012tv,Ahn:2013mva} in
the context of a similar $A_4$ model, but with
a field content quite different from ours:
we introduce only one SM singlet Majorana neutrino instead of an $A_4$ triplet, with the lepton doublets as $A_{4}$ triplets, as in Ref. \cite{He:2006dk} and many other models, but not as in Ref. \cite{Ahn:2012tv}, where they are assigned to  $A_4$ singlets.
Our scalar content is also distinct, with one additional $A_4$ triplet (and no $A_4$ singlets), which acquires a VEV in a different direction of the group space.


The  paper is organized as follows. In section \ref{model} we
outline the proposed model.
The results, in terms of neutrino masses and mixing, are presented in section \ref{massmix}.
This is followed by a numerical analysis in section \ref{Numerical Analysis}.
We conclude with discussions and a summary in \ref{Summary}. Several technical details
are presented in appendices: appendix \ref{A} collects some necessary facts about the
$A_4$ group, appendix \ref{B} contains a discussion of the full $A_4$ invariant
scalar potential, and appendix \ref{C} deals with the mass
spectrum for the physical scalars that enter in the radiative seesaw loops.

\section{The Model}
\label{model}
Our model is a multi-Higgs doublet extension of the
Standard Model (SM), with the full symmetry ${\cal G}$ experiencing a two-step spontaneous breaking
\begin{eqnarray}\label{Group}
&&{\cal G} = SU\left( 3\right) _{C}\otimes
SU\left( 2\right) _{L}\otimes U\left( 1\right) _{Y}\otimes A_{4}\otimes
Z_{2}\otimes Z^{\prime}_{2}\\
\nonumber\\[-3mm]
\nonumber
&& \hspace{35mm} \Downarrow  \Lambda_{int}\\[3mm]
\nonumber
&&  \hspace{15mm}SU\left( 3\right) _{C}\otimes
SU\left( 2\right) _{L}\otimes U\left(1\right) _{Y}\otimes Z_{2}\\[3mm]
\nonumber
&& \hspace{35mm} \Downarrow  \Lambda_{EW}\\[3mm]
\nonumber
&&  \hspace{23mm} SU\left( 3\right) _{C}\otimes U\left(1\right) _{em}\otimes Z_{2}
\end{eqnarray}
We extend the fermion sector of the SM by introducing only one
additional field, a SM singlet Majorana neutrino, $N_{R}$.  The scalar sector is significantly enlarged and contains the six
$SU(2)_{L}$ doublets $\Phi_{1,2,3}^{(1,2)}$ and three singlets $\chi_{1,2,3}$. We group them in triplets of $A_{4}$.
The complete field content and its ${\cal G}$ assignments is given below:
\begin{eqnarray}\label{FieldContent-S}
&& \Phi^{(k=1,2)}:  \left({\bf 1}, {\bf 2}, 1, {\bf 3}, (-1)^{k}, 1\right), \ \ \
\chi:  \left({\bf 1}, {\bf 1}, 1, {\bf 3}, 1, -1\right),\\
\label{FieldContent-lL}
&&l_{L}:\hspace{8.5mm} \left({\bf 1}, {\bf 2}, -1, {\bf 3}, 1, 1\right),\\
\label{FieldContent-ER}
&& e_{R}: \hspace{8.5mm}  \left({\bf 1}, \, {\bf 1}, -2, {\bf 1}, 1, 1\right), \hspace{6mm}   
\mu_{R}: \left({\bf 1}, \, {\bf 1}, -2, {\bf 1^{\prime}}, 1, 1\right), \ \ \
\tau_{R}:\left({\bf 1}, {\bf 1}, -2, {\bf 1^{\prime\prime}}, 1, 1\right), \\
\label{FieldContent-NR}
&& N_{R}: \hspace{7.5mm}  \left({\bf 1}, {\bf 1}, 0, {\bf 1}, -1, 1\right).
\end{eqnarray}
Here the numbers in bold face are dimensions of  representations of the corresponding group factor in Eq. (\ref{Group}), the third number from the left is the weak hypercharge and the last two numbers are $Z_{2}$ and $Z'_{2}$ parities, respectively.  The three families of the left-handed SM doublet leptons $l_{L}^{1,2,3}$ are unified in a single $A_{4}$ triplet $l_{L}$  while the right-handed SM singlet charged leptons  $e_{R}, \mu_{R}, \tau_{R}$
are accommodated in the three distinct $A_{4}$ singlets ${\bf 1, 1' , 1''}$. The only right-handed  SM singlet neutrino $N_{R}$ introduced in our model is assigned to ${\bf 1}$ of $A_{4}$ in order for its Majorana mass term be invariant under this symmetry. The presence of this term is crucial for our construction as explained below. Note that neither the ${\bf 1'}$ nor ${\bf 1''}$ singlet representations of $A_4$ satisfy this condition as can be seen from the multiplication rules in Eq. (\ref{A4-singlet-multiplication}).

The two SM doublet $A_4$ triplet scalars $\Phi^{(k=1,2)}$ are distinguished by their $Z_2$ parities $(-1)^{k}$.
We require that this $Z_{2}$ symmetry remains unbroken after the electroweak symmetry breaking. Therefore,  $\Phi ^{\left(1\right) }$, which transforms non-trivially under $Z_2$, does not acquire a vacuum expectation value. The preserved $Z_2$ discrete symmetry also allows for stable dark matter candidates, as in \cite{Ma:2006km,Ma:2006fn}. In our model they are either the lightest neutral component of $\Phi^{(1)}$ or the Majorana neutrino $N_{R}$. We do not address this question in the present paper.
We introduce two SM doublet $A_4$ triplets, in order to ensure that one $A_4$ scalar triplet $\Phi^{(2)}$ gives masses to the charged leptons, while the other one $\Phi^{(1)}$, with vanishing VEV, couples to the SM singlet neutrino $N_{R}$. Thus neutrinos do not receive masses at tree level.
The SM singlet $A_4$ triplet  $\chi$ is introduced in order to generate a neutrino mass matrix texture compatible with the experimentally observed deviation from the TBM pattern. As we will explain in the following, the neutrino mass matrix texture generated via the one loop seesaw mechanism is mainly due to the VEV of the SM singlet $A_4$ triplet scalar $\langle \chi\rangle = \Lambda_{int}$, which is assumed to be much larger than the scale of  the electroweak symmetry breaking $\Lambda_{int}\gg \Lambda_{EW}=246$ GeV.
In this way, the contribution associated with the $(1,1,1)$ direction in $A_{4}$-space that shapes the charged lepton mass matrix is suppressed and effectively absent in the neutrino mass matrix, leading to a mixing matrix that is TBM to a good approximation.
The $Z^{\prime}_{2}$ discrete symmetry  is also an important ingredient of our approach, as will be shown below. Once it is imposed it forbids the terms in the scalar potential involving odd powers of the SM singlet $A_4$ triplet scalar $\chi$.  This results in a reduction of the number of free model parameters and selects a particular direction of symmetry breaking in the group space. The $Z_{2}^{\prime}$ symmetry is broken after the $\chi$ field acquires a non vanishing vacuum expectation value.

With the field content of Eqs. (\ref{FieldContent-S})-(\ref{FieldContent-NR}), the Yukawa part of the model Lagrangian  for the lepton sector takes the form
\begin{equation}
\mathcal{L}_{Y}=y_{\nu }\left( \overline{l}_{L}\widetilde{\Phi }^{\left(
1\right) }\right) _{\mathbf{1}}N_{R}+M_{N}\overline{N}_{R}N_{R}^{c}+y_{e}
\left( \overline{l}_{L}\Phi ^{\left( 2\right) }\right) _{\mathbf{\bf 1}
}e_{R}+y_{\mu }\left( \overline{l}_{L}\Phi ^{\left( 2\right) }\right) _{
\mathbf{1}^{\prime \prime }}\mu _{R}+y_{\tau }\left( \overline{l}_{L}\Phi
^{\left( 2\right) }\right) _{\bf 1^{\prime }}\tau _{R}+h.c,
\label{LYlepton}
\end{equation}%
with $\widetilde{\Phi }^{\left(k \right) }= i\sigma _{2}\left( \Phi ^{\left( k\right)
}\right) ^{\ast }$ ($k=1,2$). The subscripts ${\bf 1, 1', 1''}$ denote projecting out the corresponding  $A_{4}$ singlet in the product of the two triplets.

Note that the assignment of the charged right-handed leptons (\ref{FieldContent-ER}) to different $A_{4}$ singlets leads, as can be seen in Eq. (\ref{LYlepton}),  to different Yukawa couplings $y_{e,\mu,\tau}$ of the electrically neutral components of the $\Phi^{(2)0}$ to the different charged leptons $e, \mu, \tau$.  The lightest of the $\Phi^{(2)0}$ should be interpreted  as the SM-like 125 GeV Higgs observed at the LHC
\cite{LHC-H-discovery}, and the mentioned non-universality of its couplings to the charged leptons is in agreement
with the recent ATLAS result \cite{ATLAS-CONF-2013-010}, strongly disfavoring the case of coupling universality.

As can be seen from the Appendix \ref{C}, the masses of all the neutral scalar states from the $A_4$ triplets $\Phi^{(1)}$ and $\Phi^{(2)}$,
except for the SM-like  Higgs $\Phi^{(2)0}$, are proportional to $\langle \chi\rangle = \Lambda_{int} \gg \Lambda_{EW} = 246$ GeV and consequently are very heavy. Our model is not predictive in the scalar sector, having numerous free uncorrelated parameters in the scalar potential. We  simply choose the scale  $\Lambda_{int}$  such that the heavy scalars are pushed outside the LHC reach.
The loop effects of the heavy scalars contributing to certain observables can be suppressed by the appropriate choice of the other free parameters. All these adjustments, as will be shown in Sec. \ref{Numerical Analysis},  do not affect the neutrino sector, which is totally controlled by {\it three} effective parameters, depending in turn on the scalar potential  parameters  and the lepton-Higgs Yukawa couplings.

The scalar fields $\Phi _{m}^{\left(k \right) }$ can be decomposed as:
\begin{equation}
\label{decomp}
\Phi _{m}^{\left( k \right) }=\left(
\begin{array}{c}
\frac{1}{\sqrt{2}}\left( \omega _{m}^{\left( k \right) }+i\xi _{m}^{\left(k \right) }\right) \\
\frac{1}{\sqrt{2}}\left( v_{m}^{\left(k \right) }+\rho _{m}^{\left(k \right)
}+i\eta _{m}^{\left(k \right) }\right)
\end{array}\right) ,\hspace{1cm} k = 1,2,\hspace{1cm}m=1,2,3.
\end{equation}
with
\begin{equation}
\left\langle \rho _{m}^{\left(k \right) }\right\rangle =\left\langle \eta
_{m}^{\left(k \right) }\right\rangle =\left\langle \omega _{m}^{\left(
k\right) }\right\rangle =\left\langle \xi _{m}^{\left(k \right)
}\right\rangle =0,\hspace{2cm}\hspace{35mm} k = 1,2,\hspace{1cm}m=1,2,3.
\end{equation}%
The Higgs doublets and the singlet fields can acquire vacuum expectation
values:
\begin{equation}
\left\langle \Phi _{m}^{\left( k\right) }\right\rangle =\left(
\begin{array}{c}
0 \\
\frac{v_{m}^{\left( k\right) }}{\sqrt{2}}%
\end{array}%
\right) ,\hspace{1cm}\left\langle \chi \right\rangle =\left( v_{\chi
_{1}},v_{\chi _{2}},v_{\chi _{3}}\right) ,\hspace{1cm}k=1,2,\hspace{1cm}%
m=1,2,3.
\end{equation}%
Our requirement  (see (\ref{Group})) that $Z_{2}$ is preserved implies, according to the field assignment  of (\ref{FieldContent-S}), that
\begin{equation}
\label{VEV1}
v_{m}^{\left( 1\right) }=0,\hspace{1cm}m=1,2,3.
\end{equation}%
This can be achieved by having a positive squared mass term of $\Phi^{(1)}$ in the scalar potential.
As a consequence of (\ref{LYlepton}) and (\ref{VEV1}) neutrinos do not acquire masses at tree level.
As will be discussed in more detail in section \ref{massmix}, their masses are radiatively
generated through loop diagrams involving virtual neutral scalars and the
heavy Majorana neutrino in the internal lines. The aforementioned virtual
scalars couple to real scalars due to the scalar quartic interactions, leading to the radiative seesaw mechanism of neutrino mass generation
\cite{Ma:2006km,Ma:2006fn}.

We assume the following VEV pattern for the neutral components of the  SM Higgs
doublets $\Phi _{m}^{\left( 2\right) }$ ($m=1,2,3$) and for the components
of the $A_{4}$ triplet SM singlet scalar $\chi$ :
\begin{equation}
v_{1}^{\left( 2\right) }=v_{2}^{\left( 2\right) }=v_{3}^{\left( 2\right) }=
\frac{v}{\sqrt{3}},\hspace{2cm}\left\langle \chi \right\rangle =\frac{
v_{\chi }}{\sqrt{2}}\left( 1,0,-1\right) .  \label{VEV}
\end{equation}
Here $v=\Lambda_{EW}$ and $v_{\chi} = \Lambda_{int}$.
This choice of directions in the $A_4$ space is justified by the observation that they describe a natural solution of the scalar
potential minimization equations.
Indeed, in the single-field case, $A_{4}$ invariance readily
favors
the $(1,1,1)$ direction over e.g. the $(1,0,0)$ solution for large regions of parameter space.
The vacuum $\left\langle \Phi ^{\left( 2\right) }\right\rangle $ is a
configuration that preserves a $Z_{3}$ subgroup of $A_{4}$, which has been
extensively studied by many authors (see for example Refs. \cite{Altarelli:2005yp,Altarelli:2005yx,He:2006dk,Toorop:2010ex,Ahn:2012tv,Mohapatra:2012tb,Chen:2012st}).
In our model we have more fields, but there are also classes of
the $A_{4}$ invariants favoring respective VEVs of two fields in orthogonal directions, as desired for our analysis.
Therefore our assumption is essentially that the quartic couplings in the potential
involving $\chi $ and $\Phi ^{(2)}$ are within the range of
the parameter space where these directions are the global minimum.
More details are presented in
Appendix \ref{B}, where the minimization conditions of the full
scalar potential of our model are considered, showing that the $\left\langle \chi \right\rangle $ vacuum
(\ref{VEV}), together with the $\left\langle \Phi ^{(2)}\right\rangle
$ vacuum (\ref{VEV}), are consistent.

As follows from Eqs. (\ref{LYlepton}) and (\ref{decomp}), the neutrino Yukawa interactions are described by the following Lagrangian:
\begin{equation}
\mathcal{L}_{\nu \overline{\nu }S}=\frac{y_{\nu }}{\sqrt{2}}\left[ \overline{%
\nu }_{1L}\left( \rho _{1}^{\left( 1\right) }+i\eta _{1}^{\left( 1\right)
}\right) N_{R}+\overline{\nu }_{2L}\left( \rho _{2}^{\left( 1\right) }+i\eta
_{2}^{\left( 1\right) }\right) N_{R}+\overline{\nu }_{3L}\left( \rho
_{3}^{\left( 1\right) }+i\eta _{3}^{\left( 1\right) }\right) N_{R}\right]
+h.c.
\end{equation}%
We consider the scenario where $v_{\chi }\gg v$. A moderate hierarchy in the VEVs is quite natural, given that $\chi$ is a SM
gauge singlet and its VEV does not have to be related to the electroweak scale. The scale of
$v_{\chi }$ is ultimately controlled by the $\chi $ squared mass term in the
potential.
From Eq. (\ref{LYlepton}) and Eq. (\ref{Physicalscalars})
it follows (for details see Appendix \ref{C})
that the neutrino Yukawa interactions, in terms of the physical scalar fields, can be approximately written as:
\begin{eqnarray}
\mathcal{L}_{\nu \overline{\nu }S} &=&\frac{y_{\nu }e^{i\psi}}{2}\overline{\nu }
_{1L}\left[\left( H_{1}^{0}-A_{3}^{0}\right)+i\left( H_{3}^{0}+A_{1}^{0}\right)\right] N_{R}+\frac{y_{\nu }}{\sqrt{2}}\overline{\nu }_{2L}\left(
H_{2}^{0}+iA_{2}^{0}\right) N_{R}\notag\\
&&+\frac{y_{\nu }e^{-i\psi}}{2}\overline{\nu }_{3L}\left[\left( H_{1}^{0}-A_{3}^{0}\right)+i\left( H_{3}^{0}+A_{1}^{0}\right)\right] N_{R}
+h.c.  
\label{Yukawaneutrinos}
\end{eqnarray}
When the subleading effects are considered, there is some mixing between the scalar states, so that $H_{2}^0$, $A_{2}^{0}$ will appear in the Yukawa couplings of $\overline{\nu }_{1L}$, $\overline{\nu }_{3L}$, and the other scalars will also appear in the Yukawa couplings to $\overline{\nu }_{2L}$.
As described in more detail in Appendix \ref{C}, the parameter $\psi $ is given by:
\begin{equation}
\tan 2\psi \simeq \frac{1}{\sqrt{\frac{9}{4}\left( \frac{%
M_{A_{3}^{0}}^{2}-M_{A_{1}^{0}}^{2}}{%
M_{A_{3}^{0}}^{2}+M_{A_{1}^{0}}^{2}-2M_{A_{2}^{0}}^{2}}\right) ^{2}-1}}\,,
\label{tan2psi}
\end{equation}%
in terms of the masses $M_{A_{m}^{0}}$ ($m=1,2,3$) of  the neutral CP-odd scalar fields.

\section{Lepton masses and mixing \label{massmix}}

\begin{figure}[tbh]
\vspace{-20mm}
\includegraphics[]{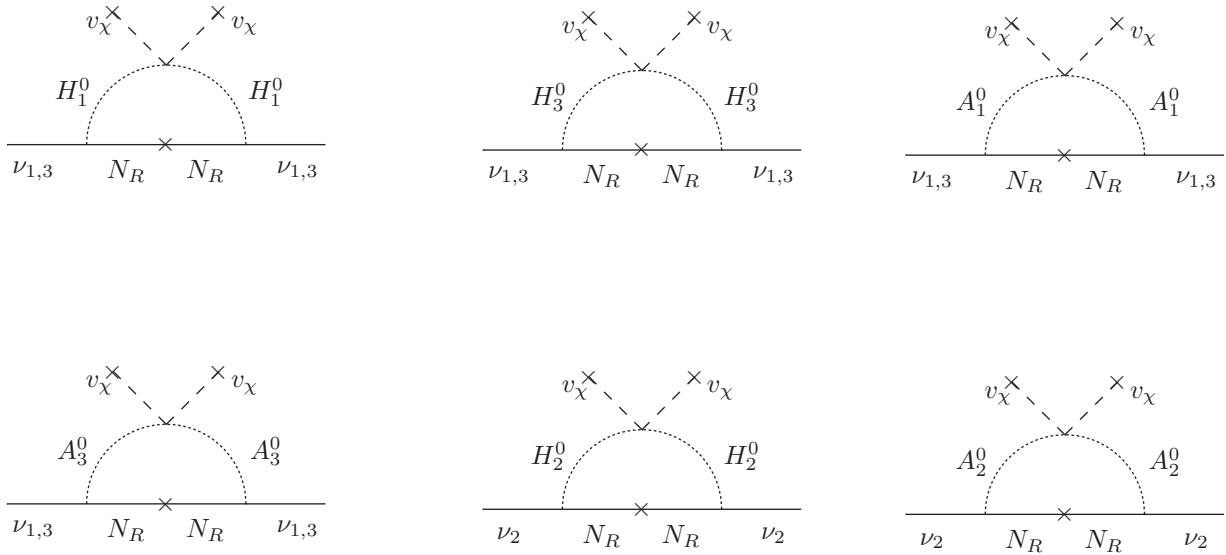}\vspace{-170mm}
\caption{One loop Feynman diagrams contributing to the entries of the neutrino mass matrix.}
\label{figMu}
\end{figure}

From Eq. (\ref{LYlepton}), and by using the product rules for the $A_{4}$ group
given in Appendix \ref{A}, it follows that the charged lepton mass matrix is
given by
\begin{equation}
M_{l}=V_{lL}^{\dag }diag\left( m_{e},m_{\mu },m_{\tau }\right) ,\hspace{2cm}
V_{lL}=\frac{1}{\sqrt{3}}\left(
\begin{array}{ccc}
1 & 1 & 1 \\
1 & \omega  & \omega ^{2} \\
1 & \omega ^{2} & \omega
\end{array}
\right) ,\hspace{2cm}\omega =e^{\frac{2\pi i}{3}}.
\label{Ml}
\end{equation}%
The
neutrino mass term does not appear at tree-level due to vanishing v.e.v. of $\Phi^{(1)}$ field (\ref{VEV1}). It arises in the form of a Majorana mass term
\begin{eqnarray}\label{Nu-Mass-Term}
&&-\frac{1}{2} \bar{\nu} M_{\nu} \nu^{C} + \mbox{h.c.}
\end{eqnarray}
from radiative corrections at 1-loop level.
The leading 1-loop contributions to the complex symmetric Majorana neutrino mass matrix $M_{\nu}$ are derived from Eqs. (\ref{Yukawaneutrinos}) and (\ref{L2}).
The corresponding diagrams are shown in  Fig. \ref{figMu}.
\vspace{3cm}
In the approximation discussed in Appendix C we obtain
\begin{equation}
M_{\nu }\simeq \left(
\begin{array}{ccc}
Ae^{2i\psi } & 0 & A \\
0 & B & 0 \\
A & 0 & Ae^{-2i\psi }%
\end{array}%
\right) .
\label{Mnu}
\end{equation}
where:
\begin{eqnarray}
\label{A-param}
A &\simeq &\frac{y_{\nu }^{2}}{16\pi ^{2}M_{N}}\left\{ \left(
M_{A_{1}^{0}}^{2}-M_{A_{2}^{0}}^{2}+\frac{\varepsilon v_{\chi }^{2}}{2}%
\right) \left[ D_{0}\left( \frac{M_{H_{1}^{0}}}{M_{N}}\right) -D_{0}\left(
\frac{M_{A_{1}^{0}}}{M_{N}}\right) \right] \right.   \notag \\
&&\hspace{14mm}+\left. \left( M_{A_{3}^{0}}^{2}-M_{A_{2}^{0}}^{2}+\frac{\varepsilon
v_{\chi }^{2}}{2}\right) \left[ D_{0}\left( \frac{M_{A_{3}^{0}}}{M_{N}}%
\right) -D_{0}\left( \frac{M_{H_{3}^{0}}}{M_{N}}\right) \right] \right\} \\[3mm],
\label{B-param}
B&\simeq& \frac{\varepsilon y_{\nu }^{2}v_{\chi }^{2}}{16\pi ^{2}M_{N}}\left[
D_{0}\left( \frac{M_{H_{2}^{0}}}{M_{N}}\right) -D_{0}\left( \frac{%
M_{A_{2}^{0}}}{M_{N}}\right) \right].
\end{eqnarray}
Here $\varepsilon $ is a dimensionless parameter, which takes into account
the difference between a pair of quartic couplings of the scalar potential
(see Appendix \ref{C} for details). We introduced the function \cite{Hernandez:2013mcf}:
\begin{equation}
D_{0}(x) =\frac{-1+x^{2}-\ln x^{2}}{\left( 1-x^{2}\right) ^{2}}.
\end{equation}
Since $M_{\nu }$ depends only on the square of the VEVs, even a moderate
hierarchy in the VEVs significantly suppresses contributions related to $%
\Phi ^{\left( 2\right) }$. Furthermore, because $\langle \chi _{2}\rangle =0$, $%
\left\vert \left( M_{\nu }\right) _{12}\right\vert $ and $\left\vert \left(
M_{\nu }\right) _{23}\right\vert $ are only generated through $\Phi ^{\left(
2\right) }$ and are then much smaller than $\left\vert \left( M_{\nu
}\right) _{13}\right\vert $ and $\left\vert \left( M_{\nu }\right)
_{mm}\right\vert $ ($m=1,2,3$).
Consequently the zero entries in Eq. (\ref{Mnu}) become
\begin{eqnarray}\label{Zero-entries}
\left( M_{\nu }\right) _{12}&\sim& \left( M_{\nu }\right) _{23}\sim \frac{v^{2}}{v_{\chi }^{2}}\left( M_{\nu }\right) _{13}
\end{eqnarray}
and are strongly suppressed in comparison to the other entries if $v_{\chi}\gg v$, as assumed in our model.
Note that a similar neutrino mass matrix texture was obtained in Ref.\cite{Memenga:2013vc} from higher dimensional operators.

The neutrino mass matrix given in Eq. (\ref{Mnu}) depends effectively only on three parameters: $A$, $B$ and $\psi$. As seen from
Eqs. (\ref{A-param}), (\ref{B-param}), the parameters $A$ and $B$ contain the dependence on various model parameters.
It is relevant that $A$ and $B$ are loop
suppressed and are approximately inverse proportional to $M_N$.
As seen from Eqs. (\ref{A-param}) and (\ref{B-param}),  a non-vanishing mass splitting between the CP even $H^{0}_{i}$ and CP odd $A^{0}_{i}$ neutral scalars is crucial. Its absence would lead to
massless neutrinos at one loop level.
Note also that universality in the quartic couplings of the scalar potential, which would correspond to $\varepsilon=0$, would imply $B\sim 0$
and lead to only one massive neutrino.
For simplicity, we parametrize the non-universality of the relevant couplings through the parameter $\varepsilon$, defined in Eq.
(\ref{univ-eps}).
As will be shown below, the Pontecorvo-Maki-Nakagawa-Sakata (PMNS) mixing matrix depends only on the parameter $\psi$,
while the neutrino mass squared splittings are controlled by parameters $A$ and $B$.

A complex symmetric Majorana mass matrix $M_{\nu}$, as in Eq. (\ref{Nu-Mass-Term}), can be diagonalized by a unitary rotation of the neutrino fields so that
\begin{eqnarray}\label{diagonalisation}
&& \nu' = V_{\nu}\cdot \nu \hspace{3mm} \longrightarrow \hspace{3mm} V^{\dagger}_{\nu} M_{\nu}(V^{\dagger}_{\nu})^T = \mbox{Diag}\{m_{\nu_{1}}, m_{\nu_{2}}, m_{\nu_{3}} \}\ \ \ \ \mbox{with} \ \ \ \ V_{\nu} V^{\dagger}_{\nu} = {\bf 1},
\end{eqnarray}
where $m_{1,2,3}$ are real and positive. The rotation matrix has the form
\begin{equation}
V_{\nu } = \left(
\begin{array}{ccc}
\cos \theta  & 0 & \sin \theta e^{-i\phi } \\
0 & 1 & 0 \\
-\sin \theta e^{i\phi } & 0 & \cos \theta
\end{array}
\right)P_{\nu},\hspace{0.5cm}\mbox{with}\hspace{0.5cm}P_{\nu}=diag\left(e^{i\alpha_1/2},e^{i\alpha_2/2},e^{i\alpha_3/2}\right),\hspace{0.5cm} \theta =\pm \frac{\pi }{4},\hspace{1cm}\phi =-2\psi .
\label{tanphi}
\end{equation}
We identify the Majorana neutrino masses and Majorana phases for the two possible solutions with $\theta = \pi/4, -\pi/4$ with the normal (NH) and inverted (IH) mass hierarchies, respectively. They are
\begin{eqnarray}\label{mass-spectrum-Normal}
\mbox{NH}&:&  \theta = + \frac{\pi}{4}: \hspace{10mm} m_{\nu_{1}} = 0, \hspace{10mm} m_{\nu_{2}} = B, \hspace{10mm} m_{\nu_{3}} = 2 A, \hspace{10mm}\alpha_1=\alpha_2=0,\hspace{10mm}\alpha_3=\phi, \\[3mm]
\label{mass-spectrum-Inverted}
\mbox{IH}&:& \theta = - \frac{\pi}{4}: \hspace{10mm} m_{\nu_{1}} = 2 A, \hspace{8mm} m_{\nu_{2}} = B, \hspace{10mm} m_{\nu_{3}} = 0, \hspace{12.5mm}\alpha_2=\alpha_3=0,\hspace{10mm}\alpha_1=-\phi.
\end{eqnarray}

Note that the nonvanishing Majorana phases are $\phi$ and $-\phi$ for normal and inverted mass hierarchies, respectively.

With the rotation matrices in the charged lepton sector $V_{lL}$, given in Eq. (\ref{Ml}), and in the neutrino sector $V_{\nu}$ , given in Eq. (\ref{tanphi}), we find
the PMNS mixing matrix:
\begin{equation}
U=V_{lL}^{\dag }V_{\nu }\simeq \left(
\begin{array}{ccc}
\frac{\cos \theta }{\sqrt{3}}-\frac{e^{i\phi }\sin \theta }{\sqrt{3}} &
\frac{1}{\sqrt{3}} & \frac{\cos \theta }{\sqrt{3}}+\frac{e^{-i\phi }\sin
\theta }{\sqrt{3}} \\
& & \\
\frac{\cos \theta }{\sqrt{3}}-\frac{e^{i\phi +\frac{2i\pi }{3}}\sin \theta }{%
\sqrt{3}} & \frac{e^{-\frac{2i\pi }{3}}}{\sqrt{3}} & \frac{e^{\frac{2i\pi }{3%
}}\cos \theta }{\sqrt{3}}+\frac{e^{-i\phi }\sin \theta }{\sqrt{3}} \\
& & \\
\frac{\cos \theta }{\sqrt{3}}-\frac{e^{i\phi -\frac{2i\pi }{3}}\sin \theta }{%
\sqrt{3}} & \frac{e^{\frac{2i\pi }{3}}}{\sqrt{3}} & \frac{e^{-\frac{2i\pi }{3%
}}\cos \theta }{\sqrt{3}}+\frac{e^{-i\phi }\sin \theta }{\sqrt{3}}%
\end{array}%
\right)P_{\nu}.  \label{PMNS}
\end{equation}%
From the standard parametrization of the leptonic mixing matrix, it follows
that the lepton mixing angles are \cite{PDG}:
\begin{eqnarray}
\label{theta-ij}
&&\sin ^{2}\theta _{12}=\frac{\left\vert U_{e2}\right\vert ^{2}}{1-\left\vert
U_{e3}\right\vert ^{2}} = \frac{1}{2\mp \cos\phi}, \hspace{20mm}
\sin ^{2}\theta _{13}=\left\vert
U_{e3}\right\vert ^{2} = \frac{1}{3}(1\pm \cos\phi), \\[3mm]
\nonumber
&&\sin ^{2}\theta _{23}=\frac{\left\vert
U_{\mu 3}\right\vert ^{2}}{1-\left\vert U_{e3}\right\vert ^{2}} = \frac{2 \mp (\cos\phi + \sqrt{3} \sin\phi)}{4 \mp 2\cos\phi},
\end{eqnarray}
where the upper sign corresponds to normal ($\theta = +\pi/4$) and the lower one to inverted ($\theta = -\pi/4$) hierarchy, respectively.
The PMNS matrix (\ref{PMNS}) of our model reproduces the magnitudes of the corresponding matrix elements of 
the TBM ansatz (\ref{TBM-ansatz}) in the limit 
$\phi =0$ and $\phi=\pi$ for the inverted and the normal hierarchy, respectively. 
In both cases the special value for $\phi$ implies that the physical neutral scalars originating from $\Phi^{(1)}$ are degenerate in mass. Notice that the lepton mixing angles are controlled by the Majorana phases $\pm\phi$, where the plus and minus signs correspond to normal and inverted mass hierarchy, respectively.

The Jarlskog invariant and the CP violating phase are given by \cite{PDG}:
\begin{equation}
J=\func{Im}\left( U_{e1}U_{\mu 2}U_{e2}^{\ast }U_{\mu 1}^{\ast }\right)\simeq-\frac{1}{6\sqrt{3}}\cos 2\theta ,\hspace{2cm}\sin \delta =\frac{8J}{\cos
\theta _{13}\sin 2\theta _{12}\sin 2\theta _{23}\sin 2\theta _{13}}.
\end{equation}%
Since $\theta =\pm \frac{\pi }{4}$, we predict $J\simeq 0$ and $\delta\simeq 0$ for $v_{\chi}\gg v$, implying that in our model CP violation is suppressed in neutrino oscillations.

\section{Phenomenological implications}
\label{Numerical Analysis}
In the following we adjust the free parameters of our model to reproduce the experimental values given in \mbox{Tables \ref{NH}, \ref{IH}} and discuss some implications of this choice of the parameters.

As seen from Eqs. (\ref{mass-spectrum-Normal}), (\ref{mass-spectrum-Inverted}) and (\ref{PMNS}), (\ref{theta-ij})
we have only {\it three} effective free parameters to fit: $\phi$, $A$ and $B$. It is noteworthy that  in our model a single parameter ($\phi$) determines all three
neutrino mixing parameters $\sin ^{2}\theta _{13}$, $\sin ^{2}\theta _{12}$
and $\sin ^{2}\theta _{23}$ as well as the Majorana phases $\alpha_{i}$.  The parameters $A$ and $B$ control the two mass squared splittings $\Delta m^{2}_{ij}$. Therefore we actually fit only $\phi$ to adjust the values of $\sin^{2}\theta _{ij}$, while $A$ and $B$ for the NH and the IH hierarchies are simply
\begin{eqnarray}\label{AB-Delta-NH}
&&\mbox{NH}:\  m_{\nu_{1}} = 0, \ \ \    m_{\nu_{2}} = B =   \sqrt{\Delta m^{2}_{21}} \approx 9 \mbox{meV}, \ \ \  m_{\nu_{3}} = 2 A = \sqrt{\Delta m^{2}_{31}} \approx 51 \mbox{meV};\\[3mm]
\label{AB-Delta-IH}
&&\mbox{IH}\hspace{2mm}  :\  m_{\nu_{2}} = B = \sqrt{\Delta m^{2}_{21} +\Delta m^{2}_{13}} \approx 50 \mbox{meV}, \ \ \ \ \  m_{\nu_{1}} = 2 A = \sqrt{\Delta m^{2}_{13}}
\approx 49 \mbox{meV}, \ \ \ m_{\nu_{3}} = 0,
\end{eqnarray}
as follows from Eqs. (\ref{mass-spectrum-Normal}), (\ref{mass-spectrum-Inverted}) and the definition $\Delta m^{2}_{ij} = m^{2}_{i}-m^{2}_{j}$.
In Eqs. (\ref{AB-Delta-NH}), (\ref{AB-Delta-IH})  we assumed the best fit values of $\Delta m^{2}_{ij}$ from the Tables \ref{NH}, \ref{IH}.

Varying the model parameter $\phi$ in Eq. (\ref{theta-ij}) we have fitted the $\sin ^{2}\theta _{ij}$  to the experimental values in Tables \ref{NH}, \ref{IH}.
The best fit result is:
\begin{eqnarray}\label{parameter-fit-NH}
&&\mbox{NH}\ :\    \phi = - 0.877\, \pi, \ \ \  \sin^{2}\theta _{12} \approx 0.34 , \ \ \ \sin^{2}\theta _{23} \approx 0.61 , \ \ \
\sin^{2}\theta _{13} \approx 0.0246; \\[3mm]
\label{parameter-fit-IH}
&&\mbox{IH}\hspace{2.5mm} :\  \phi=\ \   0.12\, \pi,   \ \ \ \ \  \sin^{2}\theta _{12} \approx 0.34 , \ \ \ \sin^{2}\theta _{23} \approx 0.6, \ \ \  \ \,
\sin^{2}\theta _{13} \approx 0.025 .
\end{eqnarray}

Comparing Eqs. (\ref{parameter-fit-NH}), (\ref{parameter-fit-IH}) with Tables \ref{NH}, \ref{IH} we see that
$\sin^{2}\theta _{13}$  and $\sin ^{2}\theta _{23}$
are in excellent agreement with the experimental data, for both NH and IH, with $\sin^{2}\theta _{12}$ within a $2\sigma $ deviation from its best fit values.

The effective parameters $A$, $B$ and $\tan\phi$ depend on various model parameters: the SM singlet neutrino Majorana mass $M_{N}$,
the quartic and bilinear couplings of the model Lagrangian (\ref{LYlepton}), (\ref{V}), as well as on the scale of $A_{4}$ symmetry breaking $v_{\chi}$.
It is worth checking that the solution in Eqs. (\ref{AB-Delta-NH})-(\ref{parameter-fit-IH}) does imply neither fine-tuning or very large values of dimensionful parameters.
For this purpose consider a point in the model parameter space with all the relevant dimensionless quartic couplings in Eqs. (\ref{V}), (\ref{univ-eps}) given by
\begin{eqnarray}\label{quartic}
\lambda = \tau_{i} = \lambda_{b1} = \alpha_{1} =  \lambda_{a1} - \varepsilon \sim 1,
\end{eqnarray}
compatible with the perturbative regime ($\lambda/4\pi < 1$). Absence of fine-tuning in this sector  favors  $\varepsilon \sim 1$.  Using  Eqs. (\ref{MH10})-(\ref{abLB}) one may derive an order of magnitude estimate
\begin{eqnarray}\label{estimate-AB}
A\sim B \sim \left(\frac{y_{\nu}}{4\pi}\right)^{2} z(\lambda, \epsilon)\  \frac{v^2_{\chi}}{M_{N}},
\end{eqnarray}
where  the function  $z(\lambda, \varepsilon) \sim 1$ for the values chosen in Eq. (\ref{quartic}).  In this estimation  we assumed
$\mu_{1}\leq v_{\chi}$ and $v_{\chi}\gg v$.     Let us also assume that the neutrino and electron Yukawa couplings in Eq. (\ref{LYlepton}) are
comparable $y_{\nu} \sim y_{e}$.  From Eqs. (\ref{LYlepton}),  (\ref{decomp}), (\ref{VEV}) and the value of the electron mass we estimate
\begin{eqnarray}\label{estimate-Yukawa}
y_{\nu} \sim y_{e} \sim 10^{-6}.
\end{eqnarray}
Then from Eqs.   (\ref{AB-Delta-NH}), (\ref{AB-Delta-IH}) and (\ref{estimate-AB}) we roughly estimate
\begin{eqnarray}\label{Fin-estimate}
m_{\nu} \sim A\sim B \sim \frac{v_{\chi}}{M_{N}} \cdot  \mbox{meV}.
\end{eqnarray}
Therefore for any value  $M_{N} \sim v_{\chi} \gg v \sim 250$ GeV  and without special tuning of the model parameters we are in the ballpark of the neutrino mass squared splittings measured in neutrino oscillation experiments (Tables \ref{NH}, \ref{IH}).  Both the scale of new physics $v_{\chi}$, related to the $A_{4}$ symmetry, and the SM singlet Majorana neutrino mass $M_{N}$ could be comparatively low, around a few TeV.

With the values of the model parameters given in Eqs. (\ref{AB-Delta-NH})-(\ref{parameter-fit-IH}), derived from the oscillation experiments, we can predict the amplitude for neutrinoless double beta ($0\nu\beta\beta$) decay, which is proportional to the effective Majorana neutrino mass
\begin{equation}
m_{\beta\beta}=\sum_jU^2_{ek}m_{\nu_k},
\label{mee}
\end{equation}
where $U^2_{ej}$ and $m_{\nu_k}$ are the PMNS mixing matrix elements and the Majorana neutrino masses, respectively.

Then, from Eqs. (\ref{tanphi})-(\ref{PMNS}) and (\ref{AB-Delta-NH})-(\ref{parameter-fit-IH}), we predict the following effective neutrino mass for both hierarchies:
\begin{eqnarray}\label{eff-mass-pred}
m_{\beta\beta}=\frac1{3}\left(B+4A\cos^2\frac{\phi}{2}\right)=
\left\{ \begin{array}{l}
 4 \  \mbox{meV}\ \ \ \ \ \ \ \mbox{for \ \ \ \ NH}\\
 50 \ \mbox{meV}\ \ \ \ \ \ \ \mbox{for \ \ \ \ IH} \\
 \end{array} \right.
\end{eqnarray}
This is beyond the reach of the present and forthcoming neutrinoless double beta decay experiments.
The presently best upper limit on this parameter $m_{\beta\beta}\leq 160$ meV comes from the recently quoted  EXO-200 experiment  \cite{Auger:2012ar}
$T^{0\nu\beta\beta}_{1/2}(^{136}{\rm Xe}) \ge 1.6 \times 10^{25}$ yr at the 90 \% CL.
This limit will be improved within the not too distant future. The GERDA
experiment \cite{Abt:2004yk,Ackermann:2012xja} is currently
moving to ``phase-II'', at the end of which  it is expected to reach \mbox{$T^{0\nu\beta\beta}_{1/2}(^{76}{\rm Ge}) \geq 2\times 10^{26}$ yr},
corresponding to $m_{\beta\beta}\leq 100$ MeV.
A bolometric CUORE experiment, using ${}^{130}Te$
\cite{Alessandria:2011rc}, is currently under construction.  Its estimated sensitivity is
around $T^{0\nu\beta\beta}_{1/2}(^{130}{\rm Te})\sim 10^{26}$ yr  corresponding to
\mbox{$m_{\beta\beta}\leq 50$ meV.}
There are also proposals for ton-scale next-to-next generation $0\nu\beta\beta$ experiments with $^{136}$Xe
\cite{KamLANDZen:2012aa,Auty:2013:zz} and $^{76}$Ge \cite{Abt:2004yk,Guiseppe:2011me} claiming
sensitivities over  $T^{0\nu\beta\beta}_{1/2} \sim 10^{27}$ yr, corresponding to
$m_{\beta\beta}\sim 12-30$ meV.  For recent experimental reviews, see for example Ref. \cite{Barabash:1209.4241} and references therein.
Thus, according to Eq. (\ref{eff-mass-pred}) our model predicts $T^{0\nu\beta\beta}_{1/2}$ at the level of sensitivities of the next generation or next-to-next generation $0\nu\beta\beta$ experiments.



\section{Conclusions}
 \label{Summary}

We have presented a simple renormalizable model that successfully accounts for the charged
lepton and neutrino masses and mixings. The neutrino masses arise from a
radiative seesaw mechanism, which explains their smallness, while
keeping the scale of new physics $\Lambda_{int}$ at the comparatively low values, which could be  about  a few TeV (for the single
SM singlet neutrino $N_{R}$).
The neutrino mixing is approximately tribimaximal due to the  spontaneously broken $A_4$ symmetry of the model.
The experimentally observed deviation from the TBM pattern is implemented by introducing the SM singlet $A_{4}$ triplet
$\chi$. Its VEV  $\langle\chi\rangle  = \Lambda_{int} \gg \Lambda_{EW}$ breaks $A_{4}$ symmetry and properly shapes the neutrino mass matrix at 1-loop level. CP violation in neutrino oscillations is suppressed.

The model has only 3 effective free parameters in the neutrino sector, which, nevertheless,  allowed us to reproduce with good accuracy
the mass squared splittings and all mixing angles measured in neutrino oscillation experiments for both normal and inverted  neutrino spectrum.

The model predicts 
the effective Majorana neutrino mass $m_{\beta\beta}$ for neutrinoless double beta decay to be 4 meV and 50 meV for the normal and the inverted neutrino spectrum, respectively.

The lightest neutral scalar of our model, $\Phi^{(2)0}$, interpreted  as the SM-like 125 GeV Higgs  boson observed at the LHC, has
non-universal Yukawa couplings to the charged leptons $e, \mu, \tau$. This is in agreement with the recent ATLAS result
\cite{ATLAS-CONF-2013-010},  strongly disfavoring the case of Yukawa coupling universality.

An unbroken $Z_2$ discrete symmetry of our model also allows for stable dark matter candidates, as in Refs.  \cite{Ma:2006km,Ma:2006fn}. The candidate could be either the lightest neutral component of $\Phi^{(1)}$ or the right-handed Majorana neutrino $N_{R}$. We do not address this possibility further in the present paper.

\section*{Acknowledgments}

This work was partially supported  by Fondecyt (Chile) under grants 1100582 and 1100287.
IdMV was supported by DFG grant PA 803/6-1 and by the Swiss National Science Foundation.
HP was supported by DFG grant PA 803/6-1.
AECH thanks Dortmund University for hospitality where part of this work was done. The visit of AECH to Dortmund University was supported by Dortmund University and DFG-CONICYT grant PA-803/7-1.

\appendix

\section{The product rules for $A_4$ \label{A}}

The following product rules for the $A_{4}$ group were used in the construction of our model Lagrangian:
\begin{eqnarray}\label{prod-rule-1}
&& \hspace{18mm }\mathbf{3}\otimes \mathbf{3}=\mathbf{3}_{s}\oplus \mathbf{3}_{a}\oplus
\mathbf{1}\oplus \mathbf{1}^{\prime }\oplus \mathbf{1}^{\prime \prime },\\[3mm]
\label{A4-singlet-multiplication}
&&\mathbf{1}\otimes \mathbf{1}=\mathbf{1},\hspace{5mm}\mathbf{1}^{\prime}\otimes \mathbf{1}^{\prime \prime }=\mathbf{1},\hspace{5mm}
\mathbf{1}^{\prime }\otimes \mathbf{1}^{\prime }=\mathbf{1}^{\prime \prime },
\hspace{5mm}\mathbf{1}^{\prime \prime }\otimes \mathbf{1}^{\prime \prime }=\mathbf{1}^{\prime },
\end{eqnarray}
Denoting $\left( x_{1},y_{1},z_{1}\right) $ and
$\left(x_{2},y_{2},z_{2}\right) $ as the basis vectors for two  $A_{4}$-triplets $\mathbf{3}$, one finds:

\begin{eqnarray}\label{triplet-vectors}
&&\left( \mathbf{3}\otimes \mathbf{3}\right) _{\mathbf{1}}=x_{1}y_{1}+x_{2}y_{2}+x_{3}y_{3},\\
&&\left( \mathbf{3}\otimes \mathbf{3}\right) _{\mathbf{3}_{s}}=\left(
x_{2}y_{3}+x_{3}y_{2},x_{3}y_{1}+x_{1}y_{3},x_{1}y_{2}+x_{2}y_{1}\right) ,
\ \ \ \
\left( \mathbf{3}\otimes \mathbf{3}\right) _{\mathbf{1}^{\prime}}=x_{1}y_{1}+\omega x_{2}y_{2}+\omega ^{2}x_{3}y_{3},\\
&&\left( \mathbf{3}\otimes \mathbf{3}\right) _{\mathbf{3}_{a}}=\left(x_{2}y_{3}-x_{3}y_{2},x_{3}y_{1}-x_{1}y_{3},x_{1}y_{2}-x_{2}y_{1}\right),
\ \ \ \left( \mathbf{3}\otimes \mathbf{3}\right) _{\mathbf{1}^{\prime\prime }}=x_{1}y_{1}+\omega ^{2}x_{2}y_{2}+\omega x_{3}y_{3},
\end{eqnarray}
where $\omega =e^{i \frac{ 2 \pi }{3}}$. The representation $\mathbf{1}$
is trivial, while the non-trivial $\mathbf{1}^{\prime }$ and $\mathbf{1}^{\prime \prime }$
are complex conjugate to each other.
Comprehensive reviews of discrete symmetries in particle physics can be found in Refs. \cite{King:2013eh,Altarelli:2010gt,Ishimori:2010au,Discret-Group-Review}.

\section{Scalar Potential \label{B}}

The scalar potential of the model is constructed of
the three  $A_{4}$ triplet fields $\Phi^{(1,2)}$ and $\chi$ in the way invariant under the group ${\cal G}$  in Eq. (\ref{Group}).

For convenience we separate its  terms into the three different groups as

\begin{equation}
V=V\left( \Phi ^{\left( 1\right) },\Phi ^{\left( 2\right) }\right) +V\left(
\Phi ^{\left( 1\right) },\Phi ^{\left( 2\right) },\chi \right) +V\left( \chi
\right) ,
\label{V}
\end{equation}
where
\begin{eqnarray}
V\left( \Phi ^{\left( 1\right) },\Phi ^{\left( 2\right) }\right)
&=&\dsum\limits_{l=1}^{2}\left[ \mu _{l}^{2}\left( \left( \Phi ^{\left(
l\right) }\right) ^{\dagger }\Phi ^{\left( l\right) }\right) _{\mathbf{1}%
}+\kappa _{l}\left( \left( \Phi ^{\left( l\right) }\right) ^{\dagger }\Phi
^{\left( l\right) }\right) _{\mathbf{1}}\left( \left( \Phi ^{\left( l\right)
}\right) ^{\dagger }\Phi ^{\left( l\right) }\right) _{\mathbf{1}}+\sigma
_{l}\left( \left( \Phi ^{\left( l\right) }\right) ^{\dagger }\Phi ^{\left(
l\right) }\right) _{\mathbf{1}^{\prime }}\left( \left( \Phi ^{\left(
l\right) }\right) ^{\dagger }\Phi ^{\left( l\right) }\right) _{\mathbf{1}%
^{\prime \prime }}\right.  \notag \\
&&+\left. \gamma _{l}\left( \left( \Phi ^{\left( l\right) }\right) ^{\dagger
}\Phi ^{\left( l\right) }\right) _{\mathbf{3s}}\left( \left( \Phi ^{\left(
l\right) }\right) ^{\dagger }\Phi ^{\left( l\right) }\right) _{\mathbf{3s}%
}+\delta _{l}\left( \left( \Phi ^{\left( l\right) }\right) ^{\dagger }\Phi
^{\left( l\right) }\right) _{\mathbf{3a}}\left( \left( \Phi ^{\left(
l\right) }\right) ^{\dagger }\Phi ^{\left( l\right) }\right) _{\mathbf{3a}}%
\right]  \notag \\
&&+\zeta _{1}\left( \left( \Phi ^{\left( 1\right) }\right) ^{\dagger }\Phi
^{\left( 1\right) }\right) _{\mathbf{3s}}\left( \left( \Phi ^{\left(
2\right) }\right) ^{\dagger }\Phi ^{\left( 2\right) }\right) _{\mathbf{3s}%
}+\zeta _{2}\left( \left( \Phi ^{\left( 1\right) }\right) ^{\dagger }\Phi
^{\left( 1\right) }\right) _{\mathbf{3a}}\left( \left( \Phi ^{\left(
2\right) }\right) ^{\dagger }\Phi ^{\left( 2\right) }\right) _{\mathbf{3a}}
\notag \\
&&+\zeta _{3}\left[ \left( \left( \Phi ^{\left( 1\right) }\right) ^{\dagger
}\Phi ^{\left( 1\right) }\right) _{\mathbf{1}^{\prime }}\left( \left( \Phi
^{\left( 2\right) }\right) ^{\dagger }\Phi ^{\left( 2\right) }\right) _{%
\mathbf{1}^{\prime \prime }}+\left( \left( \Phi ^{\left( 1\right) }\right)
^{\dagger }\Phi ^{\left( 1\right) }\right) _{\mathbf{1}^{\prime \prime
}}\left( \left( \Phi ^{\left( 2\right) }\right) ^{\dagger }\Phi ^{\left(
2\right) }\right) _{\mathbf{1}^{\prime }}\right]  \notag \\
&&+\zeta _{4}\left( \left( \Phi ^{\left( 1\right) }\right) ^{\dagger }\Phi
^{\left( 1\right) }\right) _{\mathbf{1}}\left( \left( \Phi ^{\left( 2\right)
}\right) ^{\dagger }\Phi ^{\left( 2\right) }\right) _{\mathbf{1}}+\left[
\tau _{1}\left( \left( \Phi ^{\left( 1\right) }\right) ^{\dagger }\Phi
^{\left( 2\right) }\right) _{\mathbf{3s}}\left( \left( \Phi ^{\left(
2\right) }\right) ^{\dagger }\Phi ^{\left( 1\right) }\right) _{\mathbf{3s}%
}+h.c\right]  \notag \\
&&+\left[ \tau _{2}\left( \left( \Phi ^{\left( 1\right) }\right) ^{\dagger
}\Phi ^{\left( 2\right) }\right) _{\mathbf{3a}}\left( \left( \Phi ^{\left(
2\right) }\right) ^{\dagger }\Phi ^{\left( 1\right) }\right) _{\mathbf{3a}%
}+\tau _{3}\left( \left( \Phi ^{\left( 1\right) }\right) ^{\dagger
}\Phi ^{\left( 2\right) }\right) _{\mathbf{3a}}\left( \left( \Phi ^{\left(
2\right) }\right) ^{\dagger }\Phi ^{\left( 1\right) }\right) _{\mathbf{3s}%
}+h.c\right] \notag \\
&&+\tau _{4}\left( \left( \Phi ^{\left( 1\right) }\right)
^{\dagger }\Phi ^{\left( 2\right) }\right) _{\mathbf{1}}\left( \left( \Phi
^{\left( 2\right) }\right) ^{\dagger }\Phi ^{\left( 1\right) }\right) _{%
\mathbf{1}}+\tau _{5}\left( \left( \Phi ^{\left( 1\right) }\right) ^{\dagger }\Phi
^{\left( 2\right) }\right) _{\mathbf{1}^{\prime }}\left( \left( \Phi
^{\left( 2\right) }\right) ^{\dagger }\Phi ^{\left( 1\right) }\right) _{%
\mathbf{1}^{\prime \prime }}\notag\\
&&+\tau _{6}\left( \left( \Phi ^{\left( 1\right)
}\right) ^{\dagger }\Phi ^{\left( 2\right) }\right) _{\mathbf{1}^{\prime
\prime }}\left( \left( \Phi ^{\left( 2\right) }\right) ^{\dagger }\Phi
^{\left( 1\right) }\right) _{\mathbf{1}^{\prime }},
\label{V1}
\end{eqnarray}%
\begin{eqnarray}
V\left( \Phi ^{\left( 1\right) },\Phi ^{\left( 2\right) },\chi \right)
&=&\dsum\limits_{l=1}^{2}\left\{ \lambda _{al}\left( \left( \Phi ^{\left(
l\right) }\right) ^{\dagger }\Phi ^{\left( l\right) }\right) _{\mathbf{1}%
}\left( \chi \chi \right) _{\mathbf{1}}+\lambda _{bl}\left[ \left( \left(
\Phi ^{\left( l\right) }\right) ^{\dagger }\Phi ^{\left( l\right) }\right) _{%
\mathbf{1}^{\prime }}\left( \chi \chi \right) _{\mathbf{1}^{\prime \prime
}}+\left( \left( \Phi ^{\left( l\right) }\right) ^{\dagger }\Phi ^{\left(
l\right) }\right) _{\mathbf{1}^{\prime \prime }}\left( \chi \chi \right) _{%
\mathbf{1}^{\prime }}\right] \right.\notag \\
&&+\left. \alpha _{l}\left( \left( \Phi ^{\left( l\right) }\right) ^{\dagger
}\Phi ^{\left( l\right) }\right) _{\mathbf{3s}}\left( \chi \chi \right) _{%
\mathbf{3s}}+\left[ \beta
_{l}e^{i\frac{\pi }{2}}\left( \left( \Phi ^{\left( l\right) }\right)
^{\dagger }\Phi ^{\left( l\right) }\right) _{\mathbf{3a}}\left( \chi \chi
\right) _{\mathbf{3s}}+h.c\right] \right\} ,
\label{V2}
\end{eqnarray}%
\begin{equation}
V\left( \chi \right) =D^{2}\left( \chi \chi \right) _{\mathbf{1}%
}+d_{1}\left( \chi \chi \right) _{\mathbf{1}}\left( \chi \chi \right) _{%
\mathbf{1}}+d_{2}\left( \chi \chi \right) _{\mathbf{1}^{\prime }}\left( \chi
\chi \right) _{\mathbf{1}^{\prime \prime }}+d_{3}\left( \chi \chi \right) _{%
\mathbf{3s}}\left( \chi \chi \right) _{\mathbf{3s}}.
\label{V3}
\end{equation}

Where all parameters of the scalar potential have to be real.

Now we are going to determine the conditions under which the VEV pattern for
the components of the $A_{4}$ triplet $\chi $, given in Eq. (\ref{VEV}), is a
solution of the scalar potential, assuming that the $\left\langle \Phi
^{\left( 2\right) }\right\rangle $ vacuum preserves the appropriate $Z_{3}$
subgroup of $A_4$ as in Eq. (\ref{VEV}). Then, from the previous
expressions and from the minimization conditions of the scalar potential,
the following relations are obtained:

\begin{eqnarray}
\frac{\partial V}{\partial \chi _{1}}\biggl|_{\substack{ \left\langle \chi
_{m}\right\rangle =v_{\chi _{m}}  \\ m=1,2,3}} &=&2v_{\chi _{1}}\left[ \frac{%
1}{2}\lambda _{a2}v^{2}+D^{2}+\left( 2d_{1}-d_{2}+4d_{3}\right) \left(
v_{\chi _{2}}^{2}+v_{\chi _{3}}^{2}\right) +2\left( d_{1}+d_{2}\right)
v_{\chi _{1}}^{2}\right]+\frac{2}{3}\alpha _{2}\left( v_{\chi _{2}}+v_{\chi _{3}}\right) v^{2}\notag \\
&=&0
\end{eqnarray}

\begin{eqnarray}
\frac{\partial V}{\partial \chi _{2}}\biggl|_{\substack{ \left\langle \chi
_{m}\right\rangle =v_{\chi _{m}}  \\ m=1,2,3}} &=&2v_{\chi _{2}}\left[ \frac{%
1}{2}\lambda _{a2}v^{2}+D^{2}+\left( 2d_{1}-d_{2}+4d_{3}\right) \left(
v_{\chi _{1}}^{2}+v_{\chi _{3}}^{2}\right) +2\left( d_{1}+d_{2}\right)
v_{\chi _{2}}^{2}\right] +\frac{2}{3}\alpha _{2}\left( v_{\chi _{1}}+v_{\chi _{3}}\right) v^{2}  \notag \\
&=&0
\end{eqnarray}

\begin{eqnarray}
\frac{\partial V}{\partial \chi _{3}}\biggl|_{\substack{ \left\langle \chi
_{m}\right\rangle =v_{\chi _{m}}  \\ m=1,2,3}} &=&2v_{\chi _{3}}\left[ \frac{%
1}{2}\lambda _{a2}v^{2}+D^{2}+\left( 2d_{1}-d_{2}+4d_{3}\right) \left(
v_{\chi _{1}}^{2}+v_{\chi _{2}}^{2}\right) +2\left( d_{1}+d_{2}\right)
v_{\chi _{3}}^{2}\right]+\frac{2}{3}\alpha _{2}\left( v_{\chi _{1}}+v_{\chi _{2}}\right) v^{2} \notag \\
&=&0
\end{eqnarray}

From the expressions given above, and using the vacuum configuration for the
components of the $A_{4}$ triplet $\chi $ given in Eq. (\ref{VEV}), the
following relation is obtained:
\begin{equation}
D^{2}=-\left( \frac{1}{2}\lambda_{a2}-\frac{1}{3}\alpha _{2}\right) v^{2} -\left(4d_{1}+d_{2}+4d_{3}\right) \frac{v_\chi^2}{2} \,,
\label{conditionminVEVchi}
\end{equation}
which clearly shows that the hierarchy between the VEVs depends on the $\chi \chi$ mass term ($D^2$), and that the $Z_{3}$ invariant $\left\langle \Phi^{\left( 2\right) }\right\rangle $ vacuum given in Eq. (\ref{VEV}) satisfies
the minimization conditions of the scalar potential, in a way that is consistent with the desired direction for  $\left\langle \chi \right\rangle $
This demonstrates that the VEV directions in Eq. (\ref{VEV}) are consistent with a global minimum of the scalar potential (\ref{V}) of our model, for a not fine-tuned region of parameter space.

\section{Mass spectrum of the neutral scalar fields contained in the $A_{4}$ triplet $\Phi ^{\left( 1\right) }$. \label{C}}

In this section we proceed to compute the squared mass matrix for the
neutral scalars, coming from the $A_{4}$ triplet $\Phi ^{\left( 1\right) }$.
We assume, to simplify the analysis, that the couplings are nearly universal, i.e.
\begin{eqnarray}\label{univ-eps}
\lambda =\tau_i=\lambda _{b1}=\alpha _{1}=\lambda
_{a1}-\varepsilon, \ \ \ \ \ \ \ \ \  i=(1-6).
\end{eqnarray}
In practice the coefficients do not need to be equal and indeed non-universality is required, with non-zero $\varepsilon$, necessary to generate two neutrino mass squared differences.
Using the simplified assumptions a semi-analytical treatment is possible.

As mentioned in the text, we restrict to the scenario to $v_{\chi }\gg v$.
Then, the dominant contribution to the mass Lagrangian for the neutral
scalars contained in $\Phi ^{\left( 1\right) }$ will come from $\mu
_{1}^{2}\left( \left( \Phi ^{\left( 1\right) }\right) ^{\dagger }\Phi
^{\left( 1\right) }\right) _{\mathbf{1}}+V\left( \Phi ^{\left( 1\right)
},\Phi ^{\left( 2\right) },\chi \right) $. Using the relations:
\begin{equation}
\left( \left( \Phi _{m}^{\left( 1\right) }\right) ^{\dagger }\Phi
_{m}^{\left( 1\right) }\right) =\frac{1}{2}\left[ \left( \omega _{m}^{\left(
1\right) }\right) ^{2}+\left( \xi _{m}^{\left( 1\right) }\right) ^{2}+\left(
\rho _{m}^{\left( 1\right) }\right) ^{2}+\left( \eta _{m}^{\left( 1\right)
}\right) ^{2}\right] ,\hspace{1cm}\hspace{1cm}m=1,2,3,
\end{equation}%
\begin{equation}
\left( \left( \Phi ^{\left( 1\right) }\right) ^{\dagger }\Phi ^{\left(
1\right) }\right) _{\mathbf{1}}\left\langle \left( \chi \chi \right) _{%
\mathbf{1}}\right\rangle =v_{\chi }^{2}\left[ \left( \Phi _{1}^{\left(
1\right) }\right) ^{\dagger }\Phi _{1}^{\left( 1\right) }+\left( \Phi
_{2}^{\left( 1\right) }\right) ^{\dagger }\Phi _{2}^{\left( 1\right)
}+\left( \Phi _{3}^{\left( 1\right) }\right) ^{\dagger }\Phi _{3}^{\left(
1\right) }\right] ,
\end{equation}%
\begin{equation}
\left( \left( \Phi ^{\left( 1\right) }\right) ^{\dagger }\Phi ^{\left(
1\right) }\right) _{\mathbf{1}^{\prime }}\left\langle \left( \chi \chi
\right) \right\rangle _{\mathbf{1}^{\prime \prime }}=\frac{v_{\chi }^{2}}{2}%
\left[ \left( \Phi _{1}^{\left( 1\right) }\right) ^{\dagger }\Phi
_{1}^{\left( 1\right) }+\omega \left( \Phi _{2}^{\left( 1\right) }\right)
^{\dagger }\Phi _{2}^{\left( 1\right) }+\omega ^{2}\left( \Phi _{3}^{\left(
1\right) }\right) ^{\dagger }\Phi _{3}^{\left( 1\right) }\right] \left(
1+\omega \right) ,
\end{equation}%
\begin{equation}
\left( \left( \Phi ^{\left( 1\right) }\right) ^{\dagger }\Phi ^{\left(
1\right) }\right) _{\mathbf{1}^{\prime \prime }}\left\langle \left( \chi
\chi \right) _{\mathbf{1}^{\prime }}\right\rangle =\frac{v_{\chi }^{2}}{2}%
\left[ \left( \Phi _{1}^{\left( 1\right) }\right) ^{\dagger }\Phi
_{1}^{\left( 1\right) }+\omega ^{2}\left( \Phi _{2}^{\left( 1\right)
}\right) ^{\dagger }\Phi _{2}^{\left( 1\right) }+\omega \left( \Phi
_{3}^{\left( 1\right) }\right) ^{\dagger }\Phi _{3}^{\left( 1\right) }\right]
\left( 1+\omega ^{2}\right) ,
\end{equation}%
\begin{equation}
\left( \left( \Phi ^{\left( 1\right) }\right) ^{\dagger }\Phi ^{\left(
1\right) }\right) _{\mathbf{3s}}\left\langle \left( \chi \chi \right) _{%
\mathbf{3s}}\right\rangle =-v_{\chi }^{2}\left[ \omega _{1}^{\left( 1\right)
}\omega _{3}^{\left( 1\right) }+\xi _{1}^{\left( 1\right) }\xi _{3}^{\left(
1\right) }+\rho _{1}^{\left( 1\right) }\rho _{3}^{\left( 1\right) }+\eta
_{1}^{\left( 1\right) }\eta _{3}^{\left( 1\right) }\right] ,
\end{equation}%

\begin{equation}
\beta _{1}e^{i\frac{\pi }{2}}\left( \left( \Phi ^{\left( 1\right) }\right)
^{\dagger }\Phi ^{\left( 1\right) }\right) _{\mathbf{3a}}\left\langle \left(
\chi \chi \right) _{\mathbf{3s}}\right\rangle +h.c=4\beta _{1}v_{\chi
}^{2}\left( \rho _{3}^{\left( 1\right) }\eta _{1}^{\left( 1\right) }-\rho
_{1}^{\left( 1\right) }\eta _{3}^{\left( 1\right) }\right) ,
\end{equation}

\begin{equation}
\left( \left( \Phi ^{\left( 1\right) }\right) ^{\dagger }\Phi ^{\left(
2\right) }\right) _{\mathbf{3s}}\left( \left( \Phi ^{\left( 2\right)
}\right) ^{\dagger }\Phi ^{\left( 1\right) }\right) _{\mathbf{3s}}+\left(
\left( \Phi ^{\left( 1\right) }\right) ^{\dagger }\Phi ^{\left( 2\right)
}\right) _{\mathbf{3a}}\left( \left( \Phi ^{\left( 2\right) }\right)
^{\dagger }\Phi ^{\left( 1\right) }\right) _{\mathbf{3a}}+h.c\supset \frac{%
v^{2}}{2\sqrt{3}}\left( \rho _{3}^{\left( 1\right) }\rho _{2}^{\left(
1\right) }+\rho _{1}^{\left( 1\right) }\rho _{2}^{\left( 1\right) }+\rho
_{1}^{\left( 1\right) }\rho _{3}^{\left( 1\right) }\right) ,
\end{equation}

\begin{equation}
\left( \left( \Phi ^{\left( 1\right) }\right) ^{\dagger }\Phi ^{\left(
2\right) }\right) _{\mathbf{1}}\left( \left( \Phi ^{\left( 2\right) }\right)
^{\dagger }\Phi ^{\left( 1\right) }\right) _{\mathbf{1}}\supset \frac{v^{2}}{%
4\sqrt{3}}\left( \rho _{1}^{\left( 1\right) }\rho _{1}^{\left( 1\right)
}+\rho _{2}^{\left( 1\right) }\rho _{2}^{\left( 1\right) }+\rho _{3}^{\left(
1\right) }\rho _{3}^{\left( 1\right) }+2\rho _{3}^{\left( 1\right) }\rho
_{2}^{\left( 1\right) }+2\rho _{1}^{\left( 1\right) }\rho _{2}^{\left(
1\right) }+2\rho _{1}^{\left( 1\right) }\rho _{3}^{\left( 1\right) }\right) ,
\end{equation}

\begin{eqnarray}
&&\left( \left( \Phi ^{\left( 1\right) }\right) ^{\dagger }\Phi ^{\left(
2\right) }\right) _{\mathbf{1}^{\prime }}\left( \left( \Phi ^{\left(
2\right) }\right) ^{\dagger }\Phi ^{\left( 1\right) }\right) _{\mathbf{1}%
^{\prime \prime }}+\left( \left( \Phi ^{\left( 1\right) }\right) ^{\dagger
}\Phi ^{\left( 2\right) }\right) _{\mathbf{1}^{\prime \prime }}\left( \left(
\Phi ^{\left( 2\right) }\right) ^{\dagger }\Phi ^{\left( 1\right) }\right) _{%
\mathbf{1}^{\prime }}  \notag \\
&\supset &\frac{2v^{2}}{4\sqrt{3}}\left( \rho _{1}^{\left( 1\right) }\rho
_{1}^{\left( 1\right) }+\rho _{2}^{\left( 1\right) }\rho _{2}^{\left(
1\right) }+\rho _{3}^{\left( 1\right) }\rho _{3}^{\left( 1\right) }-\rho
_{3}^{\left( 1\right) }\rho _{2}^{\left( 1\right) }-\rho _{1}^{\left(
1\right) }\rho _{2}^{\left( 1\right) }-\rho _{1}^{\left( 1\right) }\rho
_{3}^{\left( 1\right) }\right) ,
\end{eqnarray}
we obtain that the mass Lagrangian for the neutral scalars contained in $%
\Phi ^{\left( 1\right) }$ is given by:
\begin{eqnarray}
-\mathcal{L}_{mass}^{\left( 1\right) neutral} &=&\frac{\mu
_{1}^{2}+\varepsilon v_{\chi }^{2}}{2}\left[ \left( \rho _{1}^{\left(
1\right) }\right) ^{2}+\left( \rho _{2}^{\left( 1\right) }\right)
^{2}+\left( \rho _{3}^{\left( 1\right) }\right) ^{2}+\left( \eta
_{1}^{\left( 1\right) }\right) ^{2}+\left( \eta _{2}^{\left( 1\right)
}\right) ^{2}+\left( \eta _{3}^{\left( 1\right) }\right) ^{2}\right] \notag
\\
&&+\frac{3\lambda v_{\chi }^{2}}{4}\left[ \left( \rho _{1}^{\left( 1\right)
}\right) ^{2}+\left( \eta _{1}^{\left( 1\right) }\right) ^{2}+\left( \rho
_{3}^{\left( 1\right) }\right) ^{2}+\left( \eta _{3}^{\left( 1\right)
}\right) ^{2}\right]\notag \\
&&+4\beta
_{1}v_{\chi }^{2}\left( \rho _{3}^{\left( 1\right) }\eta _{1}^{\left(
1\right) }-\rho _{1}^{\left( 1\right) }\eta _{3}^{\left( 1\right) }\right) -\lambda v_{\chi }^{2}\left[ \rho _{1}^{\left( 1\right)
}\rho _{3}^{\left( 1\right) }+\eta _{1}^{\left( 1\right) }\eta _{3}^{\left(
1\right) }\right]
\notag \\
&&+\frac{\lambda v^{2}}{4\sqrt{3}}\left( 3\rho _{1}^{\left( 1\right) }\rho
_{1}^{\left( 1\right) }+3\rho _{2}^{\left( 1\right) }\rho _{2}^{\left(
1\right) }+3\rho _{3}^{\left( 1\right) }\rho _{3}^{\left( 1\right) }+2\rho
_{3}^{\left( 1\right) }\rho _{2}^{\left( 1\right) }+2\rho _{1}^{\left(
1\right) }\rho _{2}^{\left( 1\right) }+2\rho _{1}^{\left( 1\right) }\rho
_{3}^{\left( 1\right) }\right) .
\end{eqnarray}%

The squared mass matrix for the neutral scalars $(\rho_1,\rho_2,\rho_3,\eta_1,\eta_2,\eta_3)$
is given by:
\begin{equation}
M^{2}\simeq \left(
\begin{array}{cccccc}
\frac{\mu _{1}^{2}+\varepsilon v_{\chi }^{2}}{2}+\frac{3\lambda v_{\chi }^{2}%
}{4} & \frac{\lambda v^{2}}{4\sqrt{3}} & -\frac{\lambda v_{\chi }^{2}}{2} & 0
& 0 & -2\beta _{1}v_{\chi }^{2} \\
\frac{\lambda v^{2}}{4\sqrt{3}} & \frac{\mu _{1}^{2}+\varepsilon v_{\chi
}^{2}}{2} & \frac{\lambda v^{2}}{4\sqrt{3}} & 0 & 0 & 0 \\
-\frac{\lambda v_{\chi }^{2}}{2} & \frac{\lambda v^{2}}{4\sqrt{3}} & \frac{%
\mu _{1}^{2}+\varepsilon v_{\chi }^{2}}{2}+\frac{3\lambda v_{\chi }^{2}}{4}
& 2\beta _{1}v_{\chi }^{2} & 0 & 0 \\
0 & 0 & 2\beta _{1}v_{\chi }^{2} & \frac{\mu _{1}^{2}+\varepsilon v_{\chi
}^{2}}{2}+\frac{3\lambda v_{\chi }^{2}}{4} & 0 & -\frac{\lambda v_{\chi }^{2}%
}{2} \\
0 & 0 & 0 & 0 & \frac{\mu _{1}^{2}+\varepsilon v_{\chi }^{2}}{2} & 0 \\
-2\beta _{1}v_{\chi }^{2} & 0 & 0 & -\frac{\lambda v_{\chi }^{2}}{2} & 0 &
\frac{\mu _{1}^{2}+\varepsilon v_{\chi }^{2}}{2}+\frac{3\lambda v_{\chi }^{2}
}{4}
\end{array}
\right) .
\end{equation}%
Our near-universality assumption (\ref{univ-eps})  allows for an approximate analytical  diagonalization of this squared mass matrix, by a rotation matrix $R$:
\begin{equation}
R^{T}M^{2}R\simeq diag\left(
M_{H_{1}^{0}}^{2},M_{H_{2}^{0}}^{2},M_{H_{3}^{0}}^{2},M_{A_{1}^{0}}^{2},M_{A_{2}^{0}}^{2},M_{A_{3}^{0}}^{2}\right) .
\end{equation}
Due to the structure of the dominant $\chi$ VEV (\ref{VEV}), the rotation matrix mixes the 1st and 3rd components of the scalars. If we lift the universality condition on the quartic couplings, there will be subleading mixing of the 2nd component of the scalars as well. Within the near-universality approximation, the rotation matrix $R$ is:
\begin{equation}
R\simeq \left(
\begin{array}{cccccc}
\frac{\cos \psi \cos \theta _{1}}{\sqrt{2}} & -\frac{\sin \theta _{1}}{\sqrt{%
2}} & -\frac{\cos \theta _{1}\sin \psi }{\sqrt{2}} & -\frac{\sin \psi }{%
\sqrt{2}} & 0 & -\frac{\cos \psi }{\sqrt{2}} \\
\cos \psi \cos \theta _{2}\sin \theta _{1}-\sin \psi \sin \theta _{2} & \cos
\theta _{1}\cos \theta _{2} & -\cos \theta _{2}\sin \psi \sin \theta
_{1}-\cos \psi \sin \theta _{2} & 0 & 0 & 0 \\
\frac{\cos \theta _{2}\sin \psi }{\sqrt{2}}+\frac{\cos \psi \sin \theta
_{1}\sin \theta _{2}}{\sqrt{2}} & \frac{\cos \theta _{1}\sin \theta _{2}}{%
\sqrt{2}} & \frac{\cos \psi \cos \theta _{2}}{\sqrt{2}}-\frac{\sin \psi \sin
\theta _{1}\sin \theta _{2}}{\sqrt{2}} & -\frac{\cos \psi }{\sqrt{2}} & 0 &
\frac{\sin \psi }{\sqrt{2}} \\
\frac{\cos \theta _{2}\sin \psi }{\sqrt{2}}+\frac{\cos \psi \sin \theta
_{1}\sin \theta _{2}}{\sqrt{2}} & \frac{\cos \theta _{1}\sin \theta _{2}}{%
\sqrt{2}} & \frac{\cos \psi \cos \theta _{2}}{\sqrt{2}}-\frac{\sin \psi \sin
\theta _{1}\sin \theta _{2}}{\sqrt{2}} & \frac{\cos \psi }{\sqrt{2}} & 0 & -%
\frac{\sin \psi }{\sqrt{2}} \\
0 & 0 & 0 & 0 & 1 & 0 \\
\frac{\cos \psi \cos \theta _{1}}{\sqrt{2}} & -\frac{\sin \theta _{1}}{\sqrt{%
2}} & -\frac{\cos \theta _{1}\sin \psi }{\sqrt{2}} & \frac{\sin \psi }{\sqrt{%
2}} & 0 & \frac{\cos \psi }{\sqrt{2}}%
\end{array}%
\right) ,
\end{equation}%
with:
\begin{equation}
\tan 2\psi \simeq \frac{\lambda }{4\beta _{1}},\hspace{1cm}\hspace{1cm}\tan
2\theta _{1}\simeq \frac{2\lambda v^{2}}{\sqrt{6}\left( 3\lambda -8\beta
_{1}\right) v_{\chi }^{2}},\hspace{1cm}\hspace{1cm}\tan 2\theta _{2}\simeq
\frac{4\lambda ^{2}v}{\sqrt{6}\left( 9\lambda ^{2}-64\beta _{1}^{2}\right)
v_{\chi }}.
\end{equation}%
The masses of the physical neutral scalars are given by:
\begin{equation}
\label{MH10}
M_{H_{1}^{0}}^{2}\simeq M_{A_{1}^{0}}^{2}+\left\{ \left[ \frac{a^{2}+b^{2}}{4%
}+\frac{\left( a^{2}-b^{2}\right) \beta _{1}}{\sqrt{\lambda ^{2}+16\beta
_{1}^{2}}}\right] \frac{\mu _{1}^{2}+\varepsilon v_{\chi }^{2}}{v_{\chi }^{2}%
}-\frac{1}{2}ab\lambda +\frac{2ab\lambda \beta _{1}}{\sqrt{\lambda
^{2}+16\beta _{1}^{2}}}\right\} v^{2},
\end{equation}
\begin{equation}
M_{H_{2}^{0}}^{2}\simeq M_{A_{2}^{0}}^{2}+\frac{1}{4}\left[ a^{2}\left(
3\lambda -8\beta _{1}\right) +b^{2}\left( 3\lambda +8\beta _{1}\right)
+2\left( a^{2}+b^{2}\right) \frac{\mu _{1}^{2}+\varepsilon v_{\chi }^{2}}{%
v_{\chi }^{2}}+4ab\lambda \right] v^{2},
\end{equation}
\begin{equation}
M_{H_{3}^{0}}^{2}\simeq M_{A_{3}^{0}}^{2}+\left\{ \left[ \frac{a^{2}+b^{2}}{4%
}-\frac{\left( a^{2}-b^{2}\right) \beta _{1}}{\sqrt{\lambda ^{2}+16\beta
_{1}^{2}}}\right] \frac{\mu _{1}^{2}+\varepsilon v_{\chi }^{2}}{v_{\chi }^{2}%
}-\frac{1}{2}ab\lambda -\frac{2ab\lambda \beta _{1}}{\sqrt{\lambda
^{2}+16\beta _{1}^{2}}}\right\} v^{2},
\end{equation}
\begin{equation}
M_{A_{1}^{0}}^{2}\simeq \frac{1}{4}\left( 2\mu _{1}^{2}+2\varepsilon v_{\chi
}^{2}+3\lambda v_{\chi }^{2}-2\sqrt{\lambda ^{2}+16\beta _{1}^{2}}v_{\chi
}^{2}\right) ,
\end{equation}%
\begin{equation}
M_{A_{2}^{0}}^{2}\simeq \frac{\mu _{1}^{2}+\varepsilon v_{\chi }^{2}}{2},
\end{equation}%
\begin{equation}
M_{A_{3}^{0}}^{2}\simeq \frac{1}{4}\left( 2\mu _{1}^{2}+2\varepsilon v_{\chi
}^{2}+3\lambda v_{\chi }^{2}+2\sqrt{\lambda ^{2}+16\beta _{1}^{2}}v_{\chi
}^{2}\right) ,
\end{equation}
where
\begin{equation}
\label{abLB}
a\simeq \frac{\lambda }{\sqrt{6}\left( 3\lambda -8\beta _{1}\right) },%
\hspace{1cm}\hspace{1cm}b\simeq \frac{2\lambda ^{2}}{\sqrt{6}\left( 9\lambda
^{2}-64\beta _{1}^{2}\right) }.
\end{equation}
It is worth mentioning that the last five terms of Eq. (\ref{V1}), involving distinct $A_4$ invariant contractions, are responsible for the mass splitting between the CP even and CP odd neutral scalars.

From the previous expressions we obtain the relation connecting the parameter $\psi$ with the neutral scalar masses:
\begin{equation}
\tan 2\psi \simeq \frac{1}{\sqrt{\frac{9}{4}\left( \frac{%
M_{A_{3}^{0}}^{2}-M_{A_{1}^{0}}^{2}}{ %
M_{A_{3}^{0}}^{2}+M_{A_{1}^{0}}^{2}-2M_{A_{2}^{0}}^{2}}\right) ^{2}-1}}.
\end{equation}%
The physical scalars $H_{1}^{0}$, $H_{2}^{0}$, $H_{3}^{0}$, $A_{1}^{0}$, $%
A_{2}^{0}$ and $A_{3}^{0}$ are given by:
\begin{eqnarray}
&&\left(
\begin{array}{c}
H_{1}^{0} \\
H_{2}^{0} \\
H_{3}^{0} \\
A_{1}^{0} \\
A_{2}^{0} \\
A_{3}^{0}%
\end{array}%
\right) \simeq \notag \\
&\simeq &\left(
\begin{array}{cccccc}
\frac{\cos \psi \cos \theta _{1}}{\sqrt{2}} & \cos \psi \cos \theta _{2}\sin
\theta _{1}-\sin \psi \sin \theta _{2} & \frac{\cos \theta _{2}\sin \psi }{%
\sqrt{2}}+\frac{\cos \psi \sin \theta _{1}\sin \theta _{2}}{\sqrt{2}} &
\frac{\cos \theta _{2}\sin \psi }{\sqrt{2}}+\frac{\cos \psi \sin \theta
_{1}\sin \theta _{2}}{\sqrt{2}} & 0 & \frac{\cos \psi \cos \theta _{1}}{%
\sqrt{2}} \\
-\frac{\sin \theta _{1}}{\sqrt{2}} & \cos \theta _{1}\cos \theta _{2} &
\frac{\cos \theta _{1}\sin \theta _{2}}{\sqrt{2}} & \frac{\cos \theta
_{1}\sin \theta _{2}}{\sqrt{2}} & 0 & -\frac{\sin \theta _{1}}{\sqrt{2}} \\
-\frac{\cos \theta _{1}\sin \psi }{\sqrt{2}} & -\cos \theta _{2}\sin \psi
\sin \theta _{1}-\cos \psi \sin \theta _{2} & \frac{\cos \psi \cos \theta
_{2}}{\sqrt{2}}-\frac{\sin \psi \sin \theta _{1}\sin \theta _{2}}{\sqrt{2}}
& \frac{\cos \psi \cos \theta _{2}}{\sqrt{2}}-\frac{\sin \psi \sin \theta
_{1}\sin \theta _{2}}{\sqrt{2}} & 0 & -\frac{\cos \theta _{1}\sin \psi }{%
\sqrt{2}} \\
-\frac{\sin \psi }{\sqrt{2}} & 0 & -\frac{\cos \psi }{\sqrt{2}} & \frac{\cos
\psi }{\sqrt{2}} & 0 & \frac{\sin \psi }{\sqrt{2}} \\
0 & 0 & 0 & 0 & 1 & 0 \\
-\frac{\cos \psi }{\sqrt{2}} & 0 & \frac{\sin \psi }{\sqrt{2}} & -\frac{\sin
\psi }{\sqrt{2}} & 0 & \frac{\cos \psi }{\sqrt{2}}%
\end{array}%
\right) \allowbreak  \notag \\
&&\times \left(
\begin{array}{c}
\rho _{1}^{\left( 1\right) } \\
\rho _{2}^{\left( 1\right) } \\
\rho _{3}^{\left( 1\right) } \\
\eta _{1}^{\left( 1\right) } \\
\eta _{2}^{\left( 1\right) } \\
\eta _{3}^{\left( 1\right) }%
\end{array}%
\right) .  \label{Physicalscalars}
\end{eqnarray}%
After $\chi $ gets its VEV, $v_{\chi}$, and neglecting terms suppressed by powers of $v/v_{\chi }$,
the part of the Lagrangian that is obtained from the quartic
interactions with $\chi $
becomes
\begin{eqnarray}
-\mathcal{L}_{mass}^{\left( 1\right) neutral} &\supset &\left(
M_{A_{1}^{0}}^{2}-M_{A_{2}^{0}}^{2}+\frac{\varepsilon v_{\chi }^{2}}{2}%
\right) \left[ \left( H_{1}^{0}\right) ^{2}+\left( A_{1}^{0}\right) ^{2}%
\right] +\frac{\varepsilon v_{\chi }^{2}}{2}\left[ \left( H_{2}^{0}\right)
^{2}+\left( A_{2}^{0}\right) ^{2}\right]  \notag \\
&&+\left( M_{A_{3}^{0}}^{2}-M_{A_{2}^{0}}^{2}+\frac{\varepsilon v_{\chi }^{2}%
}{2}\right) \left[ \left( H_{3}^{0}\right) ^{2}+\left( A_{3}^{0}\right) ^{2}%
\right] .  \label{L2}
\end{eqnarray}

\end{document}